1

# CsI-THGEM gaseous photomultipliers for RICH and noble-liquid detectors

A. Breskin[a], V. Peskov[a,e], M. Cortesi[a], R. Budnik[a], R. Chechik[a], S. Duval[b,a], D. Thers[b], A.E.C. Coimbra[c,a], J.M.F. dos Santos[c], J.A.M. Lopes[c] C.D.R. Azevedo[d,a] and J.F.C.A. Veloso[d]

[a]*Weizmann Institute of Science, Rehovot, Israel*
[b]*Subatech, Ecole des Mines, CNRS/IN2P3 and Université de Nantes, France*
[c]*University of Coimbra, Portugal*
[d]*University of Aveiro, Portugal*
[e]*CERN, Geneva, Switzerland*

**Abstract**

The properties of UV-photon imaging detectors consisting of CsI-coated THGEM electron multipliers are summarized. New results related to detection of Cherenkov light (RICH) and scintillation photons in noble liquid are presented.

### 1. Introduction

THick Gas Electron Multipliers (THGEM) [1, 2, 3] provide efficient means of radiation detection and imagining over large area. The avalanches occur within small holes, with etched rims, mechanically drilled in metal-clad insulator plates. The THGEM electrodes, with sub-mm thickness and hole-pattern, are manufactured with standard industrial techniques. They are characterized by fast response [4], good counting-rate capability, sub-mm localization resolution [5, 6] and high attainable gains, in single- or cascaded-elements configuration, in a variety of gases, including noble ones [7, 8]. For more details on THGEMs see a recent review [3].

CsI-coated THGEM detectors (Fig. 1) were proposed for UV-photon detection in RICH [9], and were systematically investigated in view of this application [10-15]. In this article we briefly summarize our recent R&D results on CsI-THGEM photon detectors with Ne-mixtures, at normal and at cryogenic conditions, in view of their possible application in RICH and for scintillation recording in noble liquid detectors.

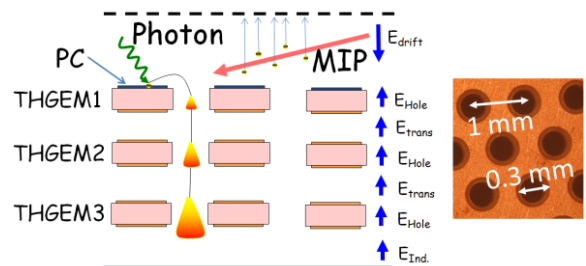

Figure 1. Schematics of a UV-photon detector with 3 cascaded THGEMs and a reflective CsI photocathode on top of the first one, and a photograph of a THGEM electrode segment.

### 2. Application in RICH

*2.1 RICH at Low particle-flux environment*

ALICE RICH with CsI-MWPCs, operating at rather low particle fluxes (currently$\leq$1Hz/mm$^2$), has been reaching gains of ~5$\times$10$^4$, with single-photon detection efficiencies of the order of 25% at 170nm [16].

In our recent studies [13] we have investigated the maximum achievable gain in CsI-THGEM photon detectors (comprising 1, 2 and 3 cascaded electrodes) and discharge probabilities in presence of low-rate radioactive sources; we used Ne/CH$_4$ and Ne/CF$_4$ of 5-10% quencher concentrations, to maintain low voltages. Comparative studies in similar experimental conditions were made with a CsI-MWPC of the geometry used in CERN-ALICE and COMPASS RICH detectors, in RICH-standard pure-CH$_4$ gas. The comparative study details are given in [13, 14]. Most important, the photoelectron extraction efficiency from the CsI-coated THGEM reached ~80% in Ne/10%CF$_4$; all the extracted photoelectrons were collected into THGEM holes, to undergo gas multiplication [11]. As detailed in [11, 13], these facts, and the high single-electron gain of the THGEM detectors, well above noise threshold [13], should result in rather high photon detection efficiencies, of about 21% with optimized THGEM-electrode geometry [11].

*2.1. RICH at high particle-flux environment*

The gain limits imposed by rate-induced discharges in the photon detectors were studied by simultaneous irradiation with UV and soft x-rays; single-, double- and triple-CsI-THGEM detectors were investigated at radiation flux of $10^3$-$10^6$ Hz/mm$^2$. It was observed that at very high flux, the gain of the CsI-THGEM dropped with rate (Fig 2), following the general trend observed in most gas-avalanche detectors [17]. E.g., a triple-CsI-THGEM (CsI-TTHGEM) operated stably at gains ~ 3-5×10$^5$ in the presence of low-intensity beta electrons mimicking MIPs (details in [13]); however, in the presence of higher-ionization x-ray background, the maximum achievable gain was significantly lower, limited by the x-ray induced total charge [13] (Fig. 2).

An important feature of CsI-THGEM is that its sensitivity to background ionization, hence the discharge rate, are significantly reduced with slightly reversed drift field above the photocathode, yet maintaining full photoelectron collection [18-20]. The discharge probably under ionizing background was indeed reduced 10-20 fold per given gain; even at high rates, the maximum attainable gain in our CsI-TTHGEM was significantly higher (about 10-fold) than in the CsI-MWPC (Fig. 2) [13]. An issue re-addressed lately was gain-stability over time, as discussed in [21, 10]. Due to considerably lower operation voltages in Ne-mixtures, the hour-scale gain stabilization-times following HV onset, are tolerable even with 0.1 mm hole-rims [6, 13].

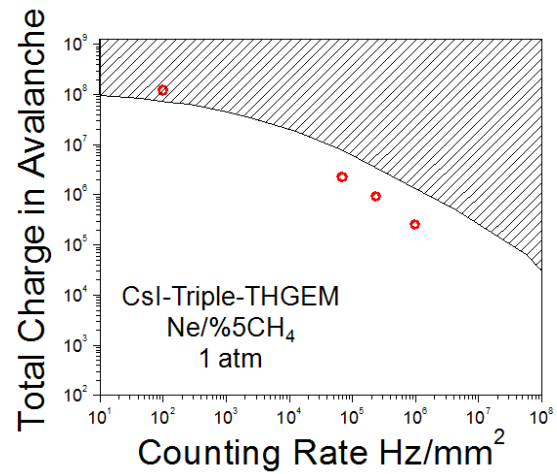

Figure 2. Maximum total avalanche charge measured in triple-THGEM with CsI photocathode (hole-diameter 0.4mm), irradiated simultaneously by UV photons and soft X-rays. Normal drift field; Ne+5%CH$_4$.The hashed "spark-region" is taken from [17].

Another important advantage of a CsI-cascaded-THGEM is the significant reduction of discharges from cathode-excitation effects, resulting from cathode bombardment by avalanche-ions [13]. As an illustration, Fig. 3 compares the rate of spurious pulses following a breakdown in CsI-THGEM and in CsI-MWPC; it clearly demonstrates that cathode excitation effect, evaluated by the rate of after-pulses, is ~100-fold weaker in CsI-THGEM. Consequently, post-discharge recovery times are very short (~30 sec), as compared to long recovery times, of up to a day, observed in CsI-MWPCs [22].

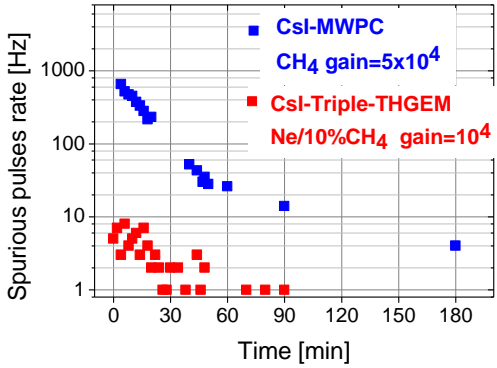

Figure 3. The rate of spurious pulses vs. time-after-discharge, in CsI-MWPC and CsI-triple-THGEM detectors; the latter has significantly reduced cathode excitation effects.

Thus our studies demonstrate that CsI-triple-THGEM, operating in Ne-based mixtures, is an attractive candidate for RICH, in both low- and high-rate background environments.

## 3. Cryogenic gaseous photon detectors

Besides the application to RICH, THGEM-photomultipliers with CsI photocathodes have been investigated in cryogenic conditions for scintillation-light recording in noble-liquid detectors. Optimized-GEM detectors (similar to THGEM) were previously proposed for the same goal [23]. Recent progress has been reported on the operation of THGEM detectors in two-phase noble-liquid detectors [24] and their readout with Geiger-mode APDs [25]. However, the implementation of gaseous photon detectors in the gas phase of noble-liquid detectors may be hindered, as the maximum gain in high-purity noble gases was found to be limited by photon-feedback effects [8]. Instead, a viable solution could be of photon detectors operating in other gas mixtures, separated by UV-windows from the noble liquid or its gas phase.

In our studies, CsI-THGEM photon-detectors were studied in Ne-mixtures at cryogenic temperatures, in view of an ongoing application in medical imaging, of a liquid-xenon (LXe) γ–camera, for a novel 3-γ medical-imaging concept [26]. Other potential applications are in light recording in large-volume noble-liquid TPCs (e.g. two-phase detectors) for Dark-matter [27] and other rare-event searches.

The latest investigations of CsI-THGEM photon detectors at cryogenic temperatures (down to ~180K), were carried out in Ne/CH$_4$ and Ne/CF$_4$ mixtures. They were carried out in a vessel partially immersed in a liquid-nitrogen/ethanol bath (Fig. 4), under gas circulation (flushed mode) or in static conditions (sealed mode). The experiments, carried out with soft x-rays and with UV photons, revealed rather high attainable gains (Fig. 5).

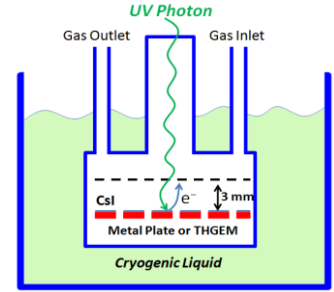

Figure 4. Experimental setup for THGEM investigations at cryogenic conditions (shown is e-extraction from single-THGEM).

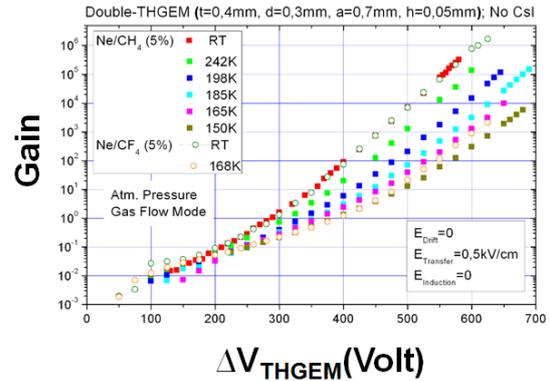

Figure 5. Absolute gain of a double-THGEM measured (current mode) in the cryogenic setup (Fig. 4) with UV photons, in Ne/CH$_4$ and Ne/CF$_4$ mixtures, at temperatures from RT down to 150K.

Photoelectron extraction studies from CsI, indicated, as expected, somewhat greater electron losses due to backscattering on the denser gas at cryogenic temperatures.

Surprisingly, at RT and pressures of 1-1.8 atm, electron-backscattering had very small dependence on gas-density, as reflected by the photocurrent plateau in Fig. 6.

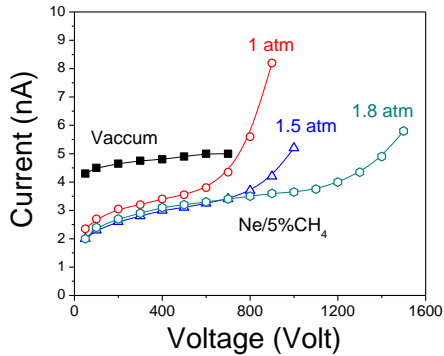

Figure 6. Photocurrent from CsI in vacuum and in Ne/5%CH$_4$.

The photoelectron extraction efficiency in gas, relative to vacuum (plateau value), is similar to that reported in [11, 28]. The cryogenic operation in gas-flow mode therefore does not foresee considerable losses in photon detection efficiencies compared to RT.

In preliminary experiments, photoelectron extraction efficiency in Ne/10%CH$_4$ dropped from 60% to 50% when cooled to ~-90°C (Fig. 7); it could be due to both, gas-density increase and residual H$_2$O condensation on CsI; it was fully restored at RT. Further R&D is in course.

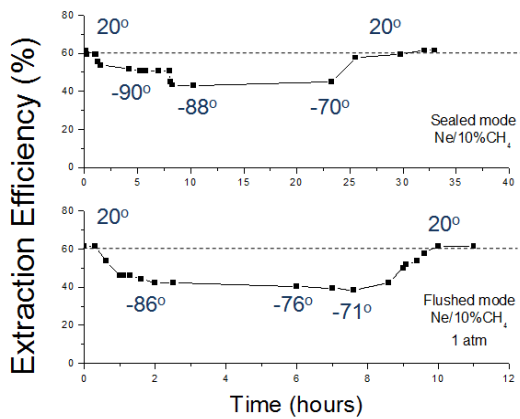

Figure 7. Preliminary results of electron-extraction efficiency vs. temperature and time, carried out in sealed and flushed modes.

## 4. Summary

- Under moderate ionizing background, a CsI-THGEM can operate at gas gains about an order of magnitude higher than a CsI-MWPC. Its maximum achievable gain is governed by the Raether limit.

- With reversed drift-filed, CsI-THGEMs can operate at high gains even in the presence of ionizing particles.

- A CsI-MWPC is more immune to heavily ionizing particles, which do not cause discharges (gain saturation); the discharges in CsI-MWPCs are due to feedback processes.

- At high counting rates CsI-MWPCs are more prone to instabilities than CsI-THGEMs due to higher probability of cathode excitation effects.

- Additionally: the photocathode's lifetime in CsI-THGEMS is considerably longer, due to reduced ion bombardment; it could be further improved with THGEMs having ion-trapping patterned electrodes [29].

- With expected photon detection efficiencies close to that of CsI-MWPC, CsI-THGEM could become suitable candidates for single-photon imaging in RICH, particularly at high-rate experiments. The choice should depend on experimental conditions.

- First, preliminary data of CsI-THGEM operation at cryogenic conditions are encouraging; more elaborate experiments are in course, in view of potential applications in Dark-matter and other rare-event experiments.

Work was partly supported by the Israel Science Foundation grant 402/05, MINERVA Foundation grant 8566, Weizmann Institute E.A. Drake & R. Drake grant and by project CERN/FP/109283/2009 under the FEDER and FCT (Lisbon) programs. V. Peskov acknowledges support of the physics faculty of Weizmann Institute. A. Breskin is the W.P. Reuther Professor of Research in The Peaceful Use of Atomic Energy.